\documentclass[prb,twocolumn,showpacs,amsfonts,amssymb,amsmath,superscriptaddress]{revtex4}

	\usepackage{bm}
	\usepackage{amsmath}
	\usepackage{amsfonts}
	\usepackage{amssymb}
	\usepackage{braket}

 	\usepackage{graphicx}

	\usepackage{hyperref}		
	\usepackage[figure]{hypcap}		

	\hypersetup{
		unicode		= false,	
		pdfborder		= {0 0 0}	
		pdftoolbar		= true,		
		pdfmenubar		= true,		
		pdffitwindow		= false,     	
		pdfstartview		= {FitH},    	
		pdfnewwindow	= true,      	
		colorlinks		= true,		
		linkcolor		= blue,		
		citecolor		= blue,		
		filecolor		= blue,		
		urlcolor		= blue,		
		linktocpage		= true,		
		pdftitle		= {},
		pdfauthor		= {M. Mucha-Kruczynski, J. Wallbank, V.I. Fal'ko},
		pdfsubject		= {},
		pdfcreator		= {MikTeX},
		pdfproducer		= {Marcin Mucha-Kruczynski},
		pdfkeywords		= {},
}

	\newcommand{\vect}[1]{\boldsymbol{#1}}
	\newcommand{\op}[1]{\hat{\boldsymbol{#1}}}
	\newcommand{\hbn}{{\it h}BN}
	\newcommand{\ghbn}{G/{\it h}BN}
\begin{document}
\title{Moir\'{e} miniband features in the angle-resolved photoemission spectra of graphene/{\hbn} heterostructures}
\author{M.~Mucha-Kruczy\'{n}ski}
\affiliation{Department of Physics, University of Bath, Claverton Down, Bath BA2~7AY, United Kingdom}
\email{m.mucha-kruczynski@bath.ac.uk}
\author{J.~R.~Wallbank}
\affiliation{National Graphene Institute, University of Manchester, Booth St E, Manchester, M13~9PL, United Kingdom}
\author{V.~I.~Fal'ko}
\affiliation{National Graphene Institute, University of Manchester, Booth St E, Manchester, M13~9PL, United Kingdom}
\affiliation{School of Physics and Astronomy, University of Manchester, Oxford Road, Manchester, M13 9PL, United Kingdom}

\begin{abstract}
We identify features in the angle-resolved photoemission spectra (ARPES) arising from the periodic pattern characteristic for graphene heterostructure with hexagonal boron nitride ({\hbn}). For this, we model ARPES spectra and intensity maps for five microscopic models used previously to describe moir\'{e} superlattice in graphene/{\hbn} systems. We show that detailed analysis of these features can be used to pin down the microscopic mechanism of the interaction between graphene and {\hbn}. We also analyse how the presence of a moir\'{e}-periodic strain in graphene or scattering of photoemitted electrons off {\hbn} can be distinguished from the miniband formation.
\end{abstract}
\pacs{73.22.Pr, 42.30.Ms, 81.05.ue}
\maketitle

\section{Introduction}

In this Article, we discuss how the moir\'{e} superlattice in graphene (G) heterostructures with hexagonal boron nitride ({\hbn}) would be reflected in angle-resolved photoemission spectroscopy (ARPES) measurements. ARPES \cite{hufner_book_2003, mahan_prb_1970} is a powerful method of exploring the electronic band structure of solids, in particular two-dimensional materials. \cite{damascelli_rmp_2003, kordyuk_ltp_2014} It was used to probe electronic states in graphene, \cite{bostwick_natphys_2007, mucha-kruczynski_prb_2008, bostwick_science_2010, siegel_pnas_2011, liu_prl_2011, hwang_prb_2011, gierz_nanoletters_2012, siegel_prl_2013} a honeycomb layer of carbon and the first of atomically thin two-dimensional atomic crystals. \cite{novoselov_science_2004} The high resolution state-of-the-art ARPES enables one to observe the modifications of the electronic dispersion in graphene due to the underlying substrate \cite{zhou_natmater_2007, enderlein_njp_2010} or in graphene grown on metal surfaces, \cite{pletikosic_prl_2009, rusponi_prl_2010, starodub_prb_2011, varykhalov_prl_2008} as well as to distinguish between Bernal stacking and twisted arrangement of layers in bilayer graphene. \cite{ohta_science_2006, ohta_prl_2012} 

In {\ghbn} van der Waals heterostructures, a difference $\delta=1.8\%$ \cite{watanabe_natmater_2004} between the lattice constants of the two crystals and a misalignment angle $\theta$ between their crystalline axes produce a quasi-periodic structure, \cite{xue_natmater_2011} known as moir\'{e} pattern with the principal period $A\approx\tfrac{a}{\sqrt{\delta^2+\theta^2}}$, where $a$ is the lattice constant of graphene. The moir\'{e} perturbation leads to the formation of minibands in the graphene electronic spectrum, revealed by STM spectra, \cite{yankowitz_natphys_2012} capacitance spectroscopy \cite{yu_natphys_2014} and transport measurements. \cite{ponomarenko_nature_2013, dean_nature_2013, hunt_science_2013} Here, we show how ARPES can help in characterising the specific details of the moir\'{e} superlattice affecting electrons in graphene, specifically, in the perfectly oriented heterostructures ($\theta=0$) which can be grown using CVD \cite{oshima_ssc_2000, usachov_prb_2010, roth_nanoletters_2013} or MBE \cite{eaves_communication} deposition of graphene on {\hbn}. We also analyze how the visibility of the distinctive miniband features is affected by the inelastic broadening of holes, which is substantial for the valence band energies where the moir\'{e} miniband structure would be most sensitive to the moire superlattice details and find that second order energy derivative of ARPES signal allows to recover the characteristic features of the ARPES maps.

\begin{figure}[b]
\centering
\includegraphics[width=1.0\columnwidth]{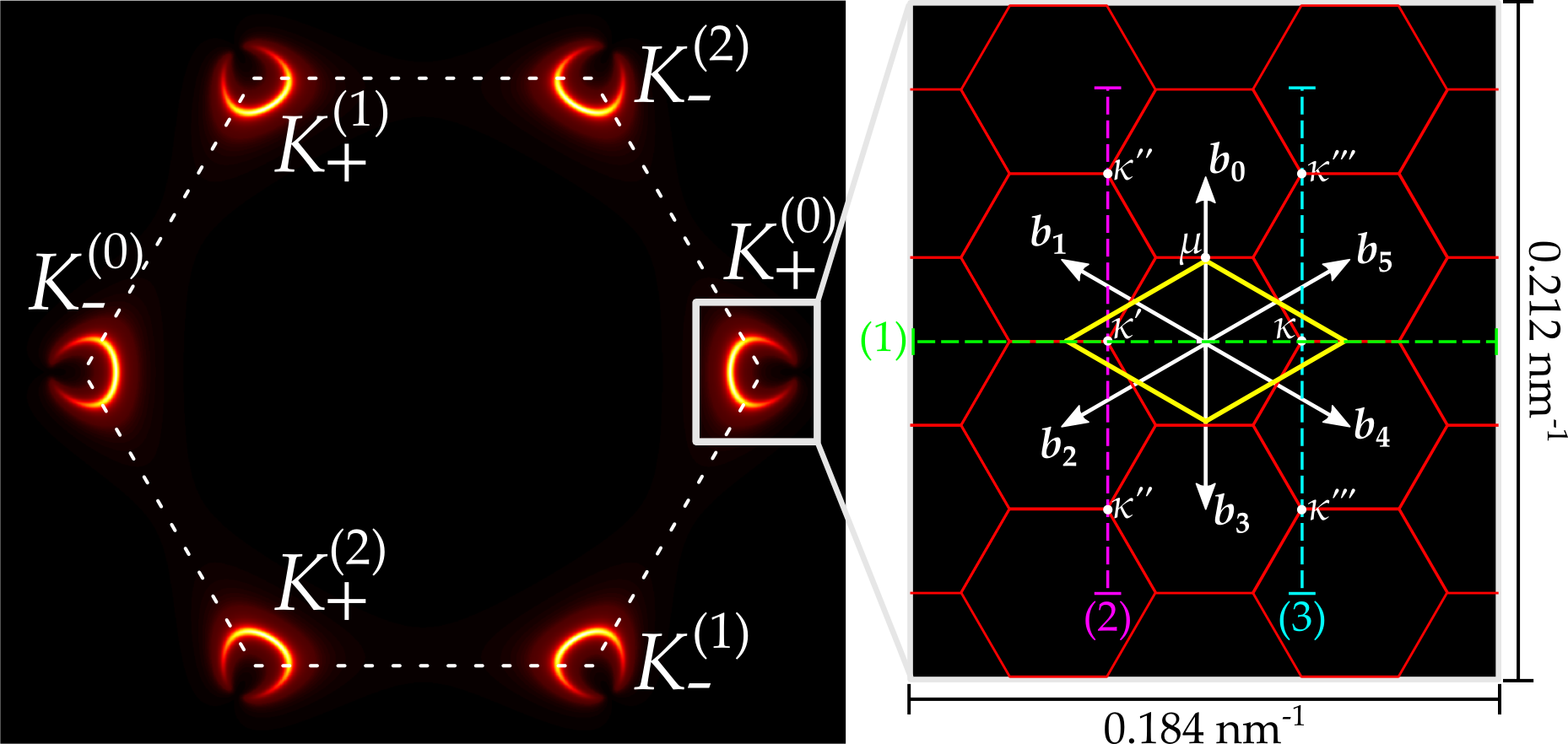}
\caption{Constant-energy ARPES map for the valence band states in free-standing graphene. White hexagon (left) depicts the Brillouin zone of graphene and the light grey rectangle the vicinity of the $K_{+}$ valley, blown up to indicate the basic reciprocal vectors of the moir\'{e} superlattice and the corresponding superlattice Brillouin zone (right). Also shown (yellow) is a rhombic primitive cell used in Fig.~\ref{fig:i}--\ref{fig:scattering}. Green, purple and cyan lines indicate cuts in the $k$-space for which ARPES spectra are presented in Figs.~\ref{fig:i}--\ref{fig:v}.}
\label{fig:introduction}
\end{figure}

The electronic bands of graphene relevant for the following study are formed by the hybridisation of $P_z$ orbitals of carbon atoms in its two triangular sublattices ($A$ and $B$). The hybridisation of $P_z$ orbitals, $\phi(\vect{r})$, on the closest lattice sites, into band states
\begin{align}\begin{split}
& \ket{\xi,\vect{p},s}_{0}\!=\!\xi\chi_{A,\xi}^{s}(\vect{p})\psi_{\vect{K}_{\xi}+\vect{p}}^{A}(\vect{r}) + \chi_{B,\xi}^{s}(\vect{p})\psi_{\vect{K}_{\xi}+\vect{p}}^{B}(\vect{r}), \\
& \psi_{\vect{k}}^{i}(\vect{r}) = \frac{1}{\sqrt{N}}\sum_{\vect{R}}e^{i\vect{k}\cdot(\vect{R}+\vect{\tau}_{i})}\phi(\vect{r}-\vect{R}-\vect{\tau}_{i}),
\label{eqn:unperturbed}
\end{split}\end{align}
produces linear dispersion, $\epsilon=sv|\vect{p}|$, of electrons near the Fermi level in undoped graphene: two cones touching with their apices exactly at the corners, $\vect{K}_{\pm}^{(j)}=\op{R}_{2\pi j/3}(\pm\tfrac{4\pi}{3a},0)^{T}$, $j=0,1,2$, of the hexagonal Brillouin zone (BZ), \cite{castro_neto_rmp_2009} see Fig.~\ref{fig:introduction}. Here, $s=1$ ($s=-1$) labels the conduction (valence) band, vectors $\vect{R}$ point to the centres of unit cells, vectors $\vect{\tau}_{A}$ and $\vect{\tau}_{B}$ indicate $A$ and $B$ sites, and $\op{R}_{\varphi}$ stands for anticlockwise rotation by angle $\varphi$. All of the corners $\vect{K}_{\xi}^{(j)}$, often called Dirac points, with the same $\xi$ are related by a graphene reciprocal vector, so that it is enough to consider in Eq.~\ref{eqn:unperturbed} only two of them, $\vect{K}_{+}^{(0)}\equiv\vect{K}_{+}$ and $\vect{K}_{-}^{(0)}\equiv\vect{K}_{-}$. Then, the coefficients $\chi_{i,\xi}^{s}(\vect{p})$ are the corresponding components of the eigenvector of the Dirac-like Hamiltonian ($\hbar=1$),
\begin{align}
\op{H}_{0} & = v\vect{p}\cdot\vect{\sigma},
\label{eqn:dirac_hamiltonian}
\end{align}
written in the basis order $\{\psi_{\vect{K}_{+}+\vect{p}}^{A},\psi_{\vect{K}_{+}+\vect{p}}^{B}\}$ in $\vect{K}_{+}$ and $\{\psi_{\vect{K}_{-}+\vect{p}}^{B},-\psi_{\vect{K}_{-}+\vect{p}}^{A}\}$ in $\vect{K}_{-}$. Pauli matrices $\sigma_{i}$, $\vect{\sigma}=(\sigma_{x},\sigma_{y})$ act in the sublattice space. Note that, as graphene is a gapless semiconductor, Fermi level in it can be easily changed by extrinsic doping.  

The ARPES intensity,
\begin{align}\begin{split}\label{eqn:intensity}
I(\vect{K}_{\xi}^{(j)}\!+\!\vect{p}_{e}) & \propto \sum_{s,\xi}\int d\vect{p}\,\left|^{s}\zeta^{\vect{K}_{\xi}^{(j)}+\vect{p}_{e}}_{\vect{K}_{\xi}+\vect{p}}\right|^{2}  \\ & \times \delta(\epsilon_{e}\!+\!W\!-\!\epsilon^{s}_{\xi,\vect{p}}\!-\!\omega), \\
^{s}\zeta^{\vect{K}_{\xi}^{(j)}+\vect{p}_{e}}_{\vect{K}_{\xi}+\vect{p}} & = \braket{e^{i(\vect{K}_{\xi}^{(j)}+\vect{p}_{e})\cdot\vect{r}}e^{ip_{e}^{\perp}z}|\xi,\vect{p},s}_{0},
\end{split}\end{align}
is determined by a projection $^{s}\zeta^{\vect{K}_{\xi}^{(j)}+\vect{p}_{e}}_{\vect{K}_{\xi}+\vect{p}}$ of an electron band state with wave vector $\vect{K}_{\xi}+\vect{p}$ onto the plane wave in vacuum with wave vector $(\vect{K}_{\xi}^{(j)}+\vect{p}_{e},p_{e}^{\perp})$, where $p_{e}^{\perp}$ denotes the out-of-plane component of photoelectron wave vector. Here, $\epsilon^{s}_{\xi,\vect{p}}$ is the initial energy of the electron in the crystal, $\omega$ is the ARPES photon energy, $W\approx 4.7$eV \cite{yu_nanoletters_2009} is the work function of graphene and $\epsilon_{e}$ is the photoelectron energy. The initial and final states are connected by the operator $\vect{A}\!\cdot\!(\vect{K}_{\xi}\!+\!\vect{p})\!\approx\!\vect{A}\!\cdot\!\vect{K}_{\xi}$, where $\vect{A}$ is the vector potential of the incoming non-circularly polarised radiation, which produces similar numerical prefactor for all $\vect{p}$.

For Dirac electrons, the relation between the two components of the electron wave-function is prescribed by the direction of electron momentum \cite{castro_neto_rmp_2009} and this electronic chirality (or pseudo-spin) provides a unique signature for graphene in the ARPES intensity. For electrons photoemitted from the $\vect{K}_{\xi}^{(j)}$ BZ corner, \cite{mucha-kruczynski_prb_2008}
\begin{align*}
I((\vect{K}_{\xi}^{(j)}+\vect{p}_{e})\sim|1+\xi se^{i\varphi_{\vect{p}_{e}}}e^{i(\vect{K}_{\xi}^{(j)}-\vect{K}_{\xi}^{(0)})\cdot\vect{d}}|^{2},
\end{align*}
where $\vect{d}=a(0,-\tfrac{1}{\sqrt{3}})$ and $\varphi_{\vect{p}}$ is the polar angle of $\vect{p}$. For the valence band, this results in crescent shapes displayed in the ARPES intensity map in Fig.~\ref{fig:introduction}, where, to take into account inelastic broadening of quasiparticles, we replaced the Dirac delta function in Eq.~\eqref{eqn:intensity} with a Lorentzian. Note that the patterns in the vicinity of the BZ corners are related to each other by $60^{\circ}$ rotations.

\section{ARPES signatures of G/\lowercase{{\it h}}BN heterostructures}

\subsection{Moir\'{e} minibands in G/\lowercase{{\it h}}BN heterostructures}

\begin{table}[b]
\centering
\caption{Values of moir\'{e} perturbation parameters in Eq.~\eqref{eqn:hamiltonian} for the microscopic models of {\ghbn} heterostructures described in the text. \cite{footnote_transformation} These parameters are dimensionless because we use unit of energy $vb$, set by the moir\'{e} pattern period, which is $vb=0.349$eV for a perfectly aligned {\ghbn} heterostructure.}
\label{table}
{
\begin{tabular}{|c|r|r|r|r|r|r|r|r|}
\hline
\multicolumn{1}{|l|}{model} & \multicolumn{1}{c|}{$u_{0}^{+}$} & \multicolumn{1}{c|}{$u_{1}^{+}$} & \multicolumn{1}{c|}{$u_{3}^{+}$} & \multicolumn{1}{c|}{$u_{0}^{-}$} & \multicolumn{1}{c|}{$u_{1}^{-}$} & \multicolumn{1}{c|}{$u_{3}^{-}$} & \multicolumn{1}{c|}{Ref.} \\ \hline
(I) & -0.0158    	& -0.1341 	& -0.0145 	& -0.0025	& 0.0081  	& 0.0086  	& \onlinecite{san-jose_prb_2014}		\\ \hline
(II) & 0.032       	& -0.063  	& -0.055   	& 0     	& 0     	& 0    	& \onlinecite{kindermann_prb_2012, wallbank_prb_2013}          	\\ \hline
(III) & -0.0241   	& -0.0191 	& -0.0134 	& -0.0097 	& 0.0087  	& 0.0089  	& \onlinecite{jung_prb_2014} 			\\ \hline
(IV) & -0.0581    	& 0.1075  	& 0.1003  	& 0.0174  	& 0.0298  	& 0.0302  	& \onlinecite{moon_prb_2014}                    	\\ \hline
(V) & -0.032	& 0.063    	& 0.055    	& 0        	& 0       	& 0        	& \onlinecite{wallbank_prb_2013}		\\ \hline
\end{tabular}
}
\end{table}

\begin{figure}[t]
\centering
\includegraphics[width=1.00\columnwidth]{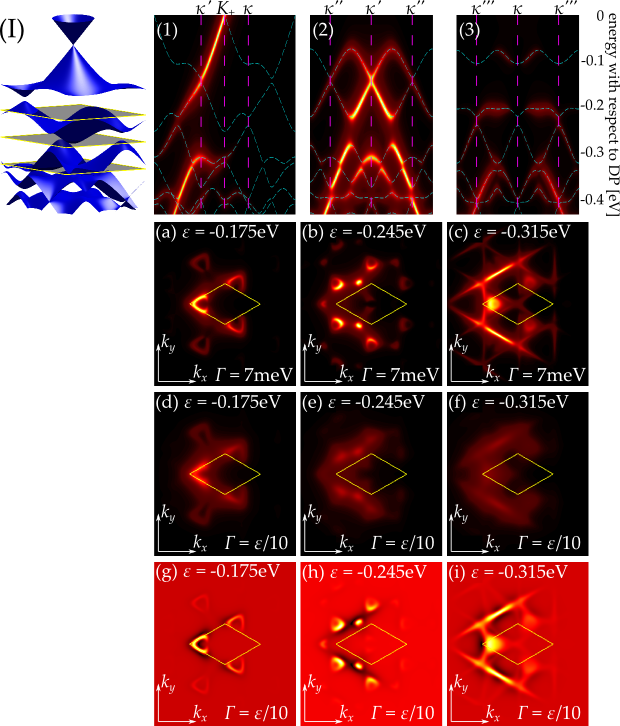}
\caption{Miniband (blue) and ARPES spectra for the model (I) as listed in Table \ref{table}. Panels (1)--(3) show spectra for cuts in the $k$-space marked in Fig.~\ref{fig:introduction}. Panels (a)--(f) display ARPES constant-energy maps for energies (a), (d) $\epsilon=-0.5vb=-0.175$eV, (b), (e) $\epsilon=-0.7vb=-0.245$eV and (c), (f) $\epsilon=-0.9vb=-0.315$eV relative to the Dirac point, corresponding to dispersion cross-sections indicated in the miniband plot. In the top row, (a)--(c), constant broadening $\Gamma=7$meV has been used while in (d)--(f) $\Gamma=\epsilon/10$. However, the same intensity scale has been used for both rows. Panels (g)--(i) show second derivative of the ARPES intensity in (d)--(f) with respect to energy. The yellow rhombus marks the sBZ and the dimensions of the maps correspond to those of the inset in the right of Fig.~\ref{fig:introduction}. }
\label{fig:i}
\end{figure}

\begin{figure}[t]
\centering
\includegraphics[width=1.00\columnwidth]{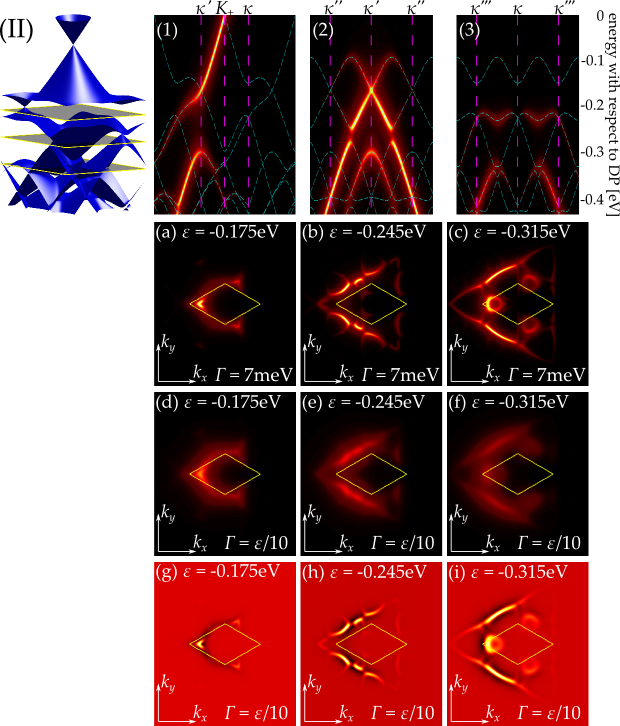}
\caption{Miniband (blue) and ARPES spectra for the model (II) as listed in Table \ref{table}. Panels (1)--(3) show spectra for cuts in the $k$-space marked in Fig.~\ref{fig:introduction}. Panels (a)--(f) display ARPES constant-energy maps for energies (a), (d) $\epsilon=-0.5vb=-0.175$eV, (b), (e) $\epsilon=-0.7vb=-0.245$eV and (c), (f) $\epsilon=-0.9vb=-0.315$eV relative to the Dirac point, corresponding to dispersion cross-sections indicated in the miniband plot. In the top row, (a)--(c), constant broadening $\Gamma=7$meV has been used while in (d)--(f) $\Gamma=\epsilon/10$. However, the same intensity scale has been used for both rows. Panels (g)--(i) show second derivative of the ARPES intensity in (d)--(f) with respect to energy. The yellow rhombus marks the sBZ and the dimensions of the maps correspond to those of the inset in the right of Fig.~\ref{fig:introduction}. }
\label{fig:ii}
\end{figure}

\begin{figure}[t]
\centering
\includegraphics[width=1.00\columnwidth]{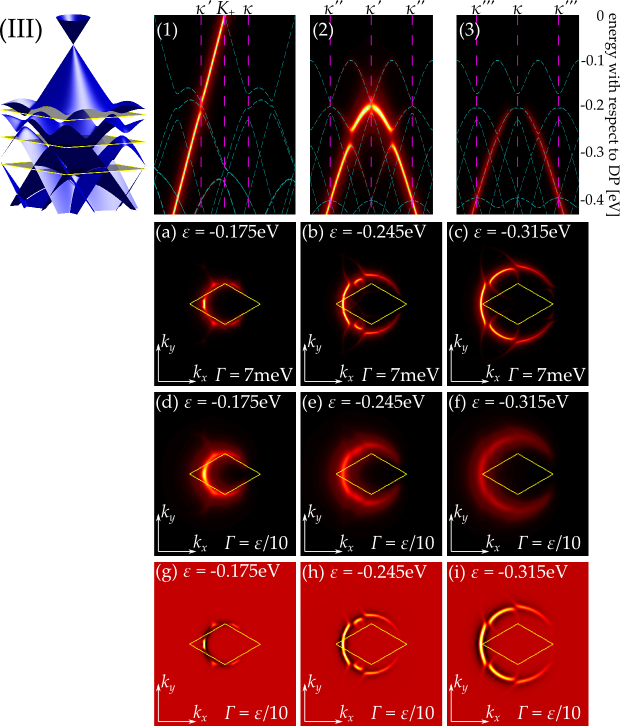}
\caption{Miniband (blue) and ARPES spectra for the model (III) as listed in Table \ref{table}. Panels (1)--(3) show spectra for cuts in the $k$-space marked in Fig.~\ref{fig:introduction}. Panels (a)--(f) display ARPES constant-energy maps for energies (a), (d) $\epsilon=-0.5vb=-0.175$eV, (b), (e) $\epsilon=-0.7vb=-0.245$eV and (c), (f) $\epsilon=-0.9vb=-0.315$eV relative to the Dirac point, corresponding to dispersion cross-sections indicated in the miniband plot. In the top row, (a)--(c), constant broadening $\Gamma=7$meV has been used while in (d)--(f) $\Gamma=\epsilon/10$. However, the same intensity scale has been used for both rows. Panels (g)--(i) show second derivative of the ARPES intensity in (d)--(f) with respect to energy. The yellow rhombus marks the sBZ and the dimensions of the maps correspond to those of the inset in the right of Fig.~\ref{fig:introduction}. }
\label{fig:iii}
\end{figure}

\begin{figure}[t]
\centering
\includegraphics[width=1.00\columnwidth]{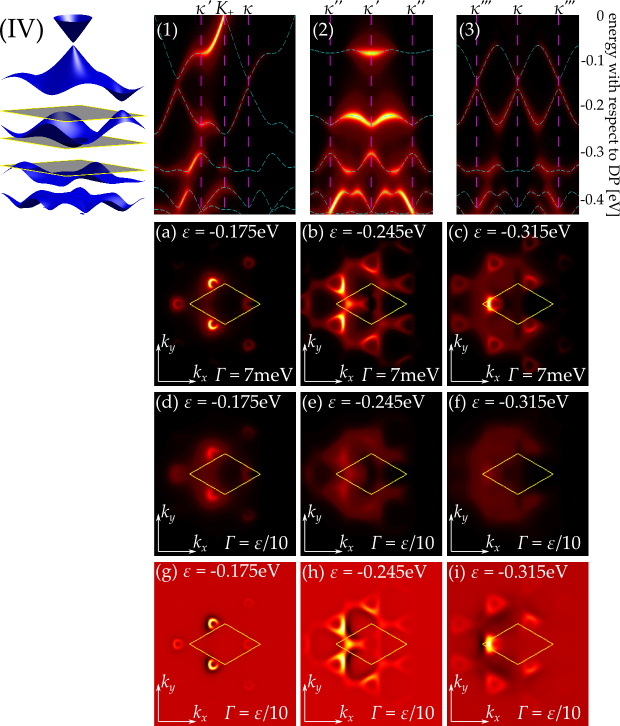}
\caption{Miniband (blue) and ARPES spectra for the model (IV) as listed in Table \ref{table}. Panels (1)--(3) show spectra for cuts in the $k$-space marked in Fig.~\ref{fig:introduction}. Panels (a)--(f) display ARPES constant-energy maps for energies (a), (d) $\epsilon=-0.5vb=-0.175$eV, (b), (e) $\epsilon=-0.7vb=-0.245$eV and (c), (f) $\epsilon=-0.9vb=-0.315$eV relative to the Dirac point, corresponding to dispersion cross-sections indicated in the miniband plot. In the top row, (a)--(c), constant broadening $\Gamma=7$meV has been used while in (d)--(f) $\Gamma=\epsilon/10$. However, the same intensity scale has been used for both rows. Panels (g)--(i) show second derivative of the ARPES intensity in (d)--(f) with respect to energy. The yellow rhombus marks the sBZ and the dimensions of the maps correspond to those of the inset in the right of Fig.~\ref{fig:introduction}. }
\label{fig:iv}
\end{figure}

\begin{figure}[t]
\centering
\includegraphics[width=1.00\columnwidth]{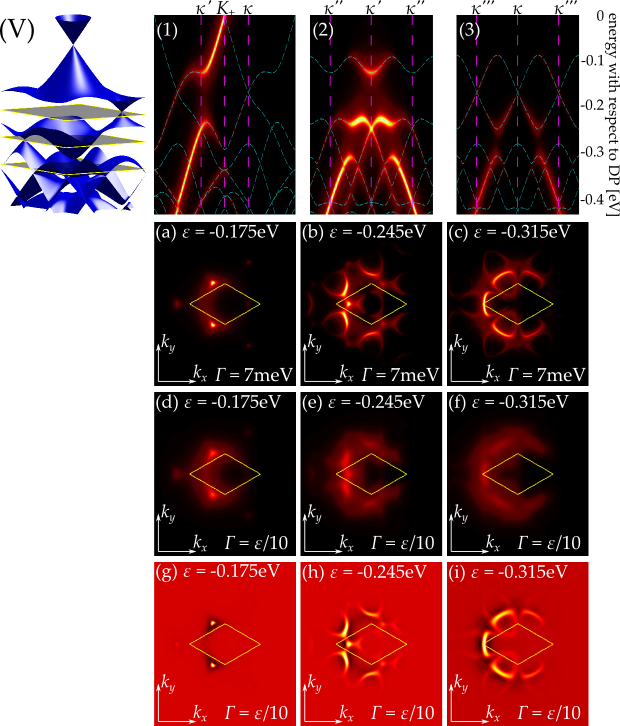}
\caption{Miniband (blue) and ARPES spectra for the model (V) as listed in Table \ref{table}. Panels (1)--(3) show spectra for cuts in the $k$-space marked in Fig.~\ref{fig:introduction}. Panels (a)--(f) display ARPES constant-energy maps for energies (a), (d) $\epsilon=-0.5vb=-0.175$eV, (b), (e) $\epsilon=-0.7vb=-0.245$eV and (c), (f) $\epsilon=-0.9vb=-0.315$eV relative to the Dirac point, corresponding to dispersion cross-sections indicated in the miniband plot. In the top row, (a)--(c), constant broadening $\Gamma=7$meV has been used while in (d)--(f) $\Gamma=\epsilon/10$. However, the same intensity scale has been used for both rows. Panels (g)--(i) show second derivative of the ARPES intensity in (d)--(f) with respect to energy. The yellow rhombus marks the sBZ and the dimensions of the maps correspond to those of the inset in the right of Fig.~\ref{fig:introduction}. }
\label{fig:v}
\end{figure}

Hexagonal boron nitride has the same honeycomb lattice as graphene, but with B and N atoms instead of carbon in the two sublattices and is a large gap ($\sim 6$eV) insulator. \cite{watanabe_natmater_2004} Placing graphene on top of {\hbn} results in a moir\'{e} superlattice that can be characterised by a Brillouin zone (sBZ) set by six basic reciprocal vectors (see inset in Fig.~\ref{fig:introduction}) $\vect{b}_{n}\!\!=\!\!\op{R}_{n\pi/3}\!\left[1\!-\!(1\!+\!\delta)^{-1}\op{R}_{\theta}\right]\!(0,\frac{4\pi}{\sqrt{3}a})$, $n\!=\!0,1,\dots,5$, where $b\!=\!\!|\vect{b}_{n}|\!\!\approx\!\!\frac{4\pi}{\sqrt{3}a}\!\sqrt{\delta^{2}\!+\!\theta^{2}}$. \cite{wallbank_prb_2013, wallbank_adp_2015} To model the electronic minibands arising due to the moir\'{e} perturbation, we use a phenomenological symmetry-based model developed in Ref.~\onlinecite{wallbank_prb_2013}. In this model, the Hamiltonian of the moir\'{e}-perturbed graphene takes a generic form suitable for all misalignment angles $\theta$
\begin{align}\begin{split}\label{eqn:hamiltonian}
\!\!\! \op{H} & \!=\! \op{H}_{0} \!+\! vb \left[ (u^{+}_{0}f_{+}+u^{-}_{0}f_{-}) \!+\! \tau_{z}\sigma_{z}(u^{+}_{3}f_{-}+u^{-}_{3}f_{+}) \right] \\ 
& + v\tau_{z}\vect{\sigma}\cdot[\vect{l_{z}}\times\nabla(u^{+}_{1}f_{-}+u^{-}_{1}f_{+})] + \Delta\tau_{z}\sigma_{z},\\
& f_{+} = \sum_{n}e^{i\vect{b_{n}}\cdot\vect{r}}, f_{-} = i\sum_{n}(-1)^{n}e^{i\vect{b_{n}}\cdot\vect{r}}, \\
\end{split}\end{align}
where the diagonal Pauli matrix $\tau_{z}$ acts in the valley space. The perturbation in Eq.~\eqref{eqn:hamiltonian} consists of a simple potential modulation, local $A$-$B$ sublattice asymmetry due to the substrate, and spatial modulation of the hopping between the $A$ and $B$ sublattices. Within each of those contributions to the moir\'{e} perturbation, the first term inside the round bracket, characterised by the dimensionless parameters $u_{i}^{+},\,i=0,1,3$, describes the inversion-symmetric part of the perturbation. Correspondingly, the second term in each round bracket, characterised by one of the dimensionless parameters $u_{i}^{-},\,i=0,1,3$, represents the inversion-asymmetric part of the perturbation. Finally, the last term describes a global gap at the Dirac point which developed due to periodic deformations in graphene with the same period as the moir\'{e} lattice. Such a global gap $\Delta\sim20$meV was used to interpret the temperature dependence of resistivity found in some heterostructures, \cite{hunt_science_2013} but in the following analysis of ARPES spectra for the states at $\epsilon\sim 50$--$400$meV from the Dirac point it will play no important role. 

The values of parameters used here to model ARPES in {\ghbn} heterostructures are listed in Table \ref{table}. \cite{footnote_transformation} They correspond to some of the microscopic models suggested for the moir\'{e} perturbation in {\ghbn} heterostructures. \cite{san-jose_prb_2014, kindermann_prb_2012, wallbank_prb_2013, jung_prb_2014, moon_prb_2014} and, for each model, the first few minibands in the valence band are displayed on the left of one of the Figs.~\ref{fig:i}--\ref{fig:v}. Models (I) \cite{san-jose_prb_2014} and (II) \cite{kindermann_prb_2012, wallbank_prb_2013} consider interlayer G--{\hbn} hopping, with (I) allowing for a periodic lattice deformation to minimise van der Waals interaction between carbon atoms and nitrogens/borons. Model (III) \cite{jung_prb_2014} is based on DFT calculations and also takes into account relaxation of the graphene lattice on top of {\hbn}. Model (IV) \cite{moon_prb_2014} uses Slater-Koster-type approach to calculate electron hopping between atomic sites within the tight-binding approximation. Finally, model (V) \cite{wallbank_prb_2013} assumes that the perturbation is caused by quadrupole electric moments placed on the atomic sites of {\hbn}. All of the five selected models predict a single secondary Dirac point (sDP) between the first and second miniband on the valence-band side located at either $\kappa'$ [models (I)-(III)] or $\kappa$ [models (IV) and (V)]. Models (I), (III) and (IV) contain inversion-asymmetric terms and hence display gaps (most pronounced for model (IV) \cite{abergel_njp_2013, chen_prb_2014}) at the sDP. The generic properties of the miniband spectra produced by all the models agree with the transport and magnetocapacitance data taken on various {\ghbn} heterostructures. \cite{yu_natphys_2014, ponomarenko_nature_2013, dean_nature_2013, hunt_science_2013} Also, optical absorption data in Ref.~\onlinecite{shi_natphys_2014} agree with spectral properties of model (II) and (V). \cite{abergel_njp_2013, abergel_prb_2015} The parameter sets we chose correspond to a perfect alignment of graphene and {\hbn} ($\theta=0$), what should be the case in CVD \cite{oshima_ssc_2000, usachov_prb_2010, roth_nanoletters_2013} or MBE-grown \cite{eaves_communication} {\ghbn} heterostructures, more relevant for ARPES studies than exfoliated graphene. In the case of such an aligned heterostructure, the characteristic energy $vb=0.349$eV.

\subsection{Minibands signature in ARPES}

The eigenstates of the superlattice Hamiltonian \eqref{eqn:hamiltonian},
\begin{align*}
\ket{\xi,\vect{p},\{m,s\}} = \sum_{\vect{g}=n_{1}\vect{b}_{1}+n_{2}\vect{b}_{2}}\sum_{s}c_{\vect{g},s}^{m,\xi}(\vect{p})\ket{\xi,\vect{g}+\vect{p},s}_{0},
\end{align*}
where $\vect{p}\in\mathrm{sBZ}$ and $m$ labels the minibands on the conduction/valence side, are the result of Bragg scattering of Dirac electrons by the moir\'{e} perturbation. We find the coefficients $c_{\vect{g},s}^{m,\xi}(\vect{p})$ and the corresponding miniband energy for electrons, $\epsilon^{\{m,s\}}_{\xi,\vect{p}}$, numerically, and, then, use those to evaluate ARPES intensity,
\begin{align}\label{eqn:arpes_intensity}
& I(\vect{K}_{\xi}^{(j)}\!+\vect{p}_{e}) \propto \sum_{m}\left| \sum_{s}c_{\vect{g},s}^{m,\xi}(\vect{p}_{e}-\vect{g}) \right.\\ \times & \left.\left[ 1+se^{i\varphi_{\vect{p}_{e}}}e^{i(\vect{K}_{\xi}^{(j)}-\vect{K}_{\xi}^{(0)})\cdot\vect{d}} \right] \right|^{2} \delta(\epsilon_{e}+W-\epsilon^{\{m,s\}}_{\xi,\vect{p}_{e}-\vect{g}}-\omega).\nonumber
\end{align} 
Here, $\vect{g}$ is the moir\'{e} reciprocal vector that brings $\vect{p}_{e}$ into the sBZ. 

We show the ARPES dispersion cuts and intensity maps for all the five models listed in Table \ref{table} in Figs.~\ref{fig:i}--\ref{fig:v}. In the top row, next to the miniband spectra, we present the ARPES images of dispersion cuts along the following $k$-space directions: (1) $\vect{k}=(k,0)$; (2) $\vect{k}=(-b/\sqrt{3},k)$; (3) $\vect{k}=(b/\sqrt{3},k)$, displayed in green, purple and cyan in the inset of Fig.~\ref{fig:introduction}. As before, we replaced the Dirac delta function with a Lorentzian and for those cuts used half-width at half-maximum $\Gamma=0.02vb=7$meV. In all the cases, the spectra display gaps and deviations from the linear traces observed for unperturbed graphene, which indicate boundaries of the sBZ. The cuts (1) for all the models except (III) and (IV) are qualitatively similar, despite different locations of the sDPs in the miniband spectra. This is because of the chirality-induced supression of signal for momentum states with $k_{x}>0$ [compare the relative intensity of cuts (2) and (3)] and small magnitude of any potential gap at the sDP. As a result, the large gap visible in cuts (1) at $\kappa'$ can be either between the first and second miniband [models (IV) and (V)] or the second and third [models (I) and (II)]. A combination of cuts along $k_{x}$ and $k_{y}$ for negative $k_{x}$ is necessary to deduce the position of the sDP accurately. 

We also show in Figs.~\ref{fig:i}--\ref{fig:v} ARPES intensity maps for energies $\epsilon=-0.5vb=-0.175$eV, $\epsilon=-0.7vb=-0.245$eV and $\epsilon=-0.9vb=-0.315$eV in the valence band, counted from the Dirac point and corresponding to miniband dispersion surfaces as indicated by the grey planes cutting through the miniband spectra. For the panels (a)--(c) of each of the Figs.~\ref{fig:i}--\ref{fig:v}, we used $\Gamma=0.02vb=7$meV, whereas for panels (d)--(f), $\Gamma=0.1/\epsilon$, to model experimental broadening as measured in Ref.~\onlinecite{siegel_pnas_2011}. In all the images, the chirality of graphene electrons is responsible for the modulation of the intensity as a function of the polar angle $\varphi$ of the electron momentum, akin to the spectra of unperturbed graphene. Similar modulation of the intensity is also clear for features in the vicinity of the sDPs, see for example Figs.~\ref{fig:i}(a), \ref{fig:ii}(c) or \ref{fig:iv}(a). The moir\'{e} effect is least pronounced for model (III), characterised by the smallest amplitudes of the moir\'{e} perturbation parameters. Maps for all the other models show strongly triangularly deformed shapes. Increased broadening used in panels (d)--(f) washes out distinctive features of the spectra. In particular, the maps for energy $\epsilon=-0.315$eV, panel (f), show only a blurred crescent-like shape resembling that of unperturbed graphene. However, the ARPES maps can be sharpened by differentiating the ARPES signal twice with respect to energy. \cite{zhang_rsi_2011} Results of such a procedure are shown in the last row in Figs.~\ref{fig:i}--\ref{fig:v}, panels (g)--(i), where we display the maps of the second order energy derivative of the ARPES intensity calculated for $\Gamma=0.1/\epsilon$. Despite this significant broadening as compared to panels (a)--(c), the characteristic features are similar. Note, however, that second derivative introduces certain spurious features in the vicinity of intensity peaks.    

\begin{figure}[b]
\centering
\includegraphics[width=\columnwidth]{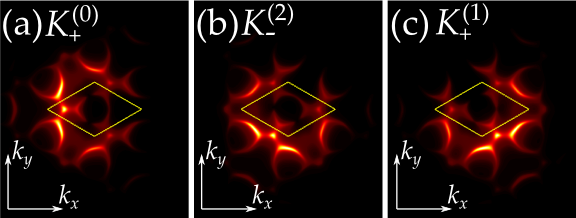}
\caption{ARPES constant-energy maps at energy $\epsilon=-0.7vb=-0.245$eV for model (V) from Table \ref{table} for different BZ corners (a) $\vect{K}_{+}^{(0)}$, (b) $\vect{K}_{-}^{(2)}$ and (c) $\vect{K}_{+}^{(1)}$. The yellow rhombus in each panel shows the sBZ boundary and the dimensions of the maps correspond to those of the inset in the right of Fig.~\ref{fig:introduction}.}
\label{fig:minibands_rotation}
\end{figure}

The time-inversion symmetry which connects the electronic states at $\vect{K}_{+}$ and $\vect{K}_{-}$ and the phase factor $e^{i(\vect{K}_{\xi}^{(j)}-\vect{K}_{\xi}^{(0)})\cdot\vect{d}}$ in Eq.~\eqref{eqn:arpes_intensity} guarantee that the ARPES patterns in consecutive BZ corners are related by $60^{\circ}$ rotation. Examples of ARPES maps in the vicinity of BZ corners $\vect{K}_{+}^{(0)}$, $\vect{K}_{-}^{(2)}$ and $\vect{K}_{+}^{(1)}$ have been shown for model (V) and energy $\epsilon=-0.245$eV in Fig.~\ref{fig:minibands_rotation}. 

\section{ARPES signature of periodic deformation pattern in G/\lowercase{{\it h}}BN heterostructures}

As noticed in Ref.~\onlinecite{woods_natphys_2014}, the graphene lattice may periodically deform to adjust locally to the slightly incommensurate {\hbn} substrate. In addition to affecting the moir\'{e} perturbation [an effect already included in models (I) and (III)], these deformations will further modify the ARPES intensity maps by altering the positions of the carbon atoms from which electrons are emitted. For a smooth deformation, the shift of atomic positions 
\begin{align*}
\vect{R}+\vect{\tau}_{i}\rightarrow\vect{R}+\vect{\tau}_{i}+\vect{u}(\vect{R}+\vect{\tau}_{i})\approx\vect{R}+\vect{\tau}_{i}+\vect{u}(\vect{R}),
\end{align*}
where $\vect{u}(\vect{r})$ is the deformation field, \cite{strain_footnote} leads to additional phases in the crystal wave function projections onto plane waves in vacuum,
\begin{align}\label{eqn:strain1}
& ^{s}\zeta^{\vect{K}_{\xi}^{(j)}+\vect{p}_{e}}_{\vect{K}_{\xi}+\vect{p}} = \left( \sum_{\vect{R}}e^{i(\vect{K}_{\xi}^{(0)}-\vect{K}_{\xi}^{(j)}+\vect{p}-\vect{p}_{e})\cdot[\vect{R}+\vect{u}(\vect{R})]} \right) \\ & \times \!\left[ \chi^{s}_{A,\xi}e^{i(\vect{K}_{\xi}-\vect{K}_{\xi}^{(j)}+\vect{p}-\vect{p}_{e})\cdot\vect{\tau}_{A}} + \chi^{s}_{B,\xi}e^{i(\vect{K}_{\xi}-\vect{K}_{\xi}^{(j)}+\vect{p}-\vect{p}_{e})\cdot\vect{\tau}_{B}} \right].\nonumber
\end{align}
We have $\vect{p},\vect{p}_{e}\ll\tfrac{1}{a}$, so that in the vicinity of the BZ corners $\vect{K}_{\xi}^{(0)}$ we can expand to first order in $(\vect{p}-\vect{p}_{e})\cdot\vect{u}(\vect{R})\ll 1$. We also use the periodicity of the deformation field and rewrite it as the Fourier transform $\vect{u}(\vect{R})=\sum_{n}\vect{u}_{n}\exp(i\vect{b}_{n}\cdot\vect{R})$, where we assume that only the simplest harmonics are important, to obtain
\begin{align*}\begin{split}
& \sum_{\vect{R}}e^{i(\vect{p}-\vect{p}_{e})\cdot[\vect{R}+\vect{u}(\vect{R})]} \approx \sum_{\vect{R}}e^{i(\vect{p}-\vect{p}_{e})\cdot\vect{R}} \\ &+ \sum_{\vect{R}}\sum_{n}i(\vect{p}-\vect{p}_{e})\cdot\vect{u}_{n}e^{i(\vect{p}-\vect{p}_{e}+\vect{b}_{n})\cdot\vect{R}}. 
\end{split}\end{align*}
\begin{figure}[t]
\centering
\includegraphics[width=\columnwidth]{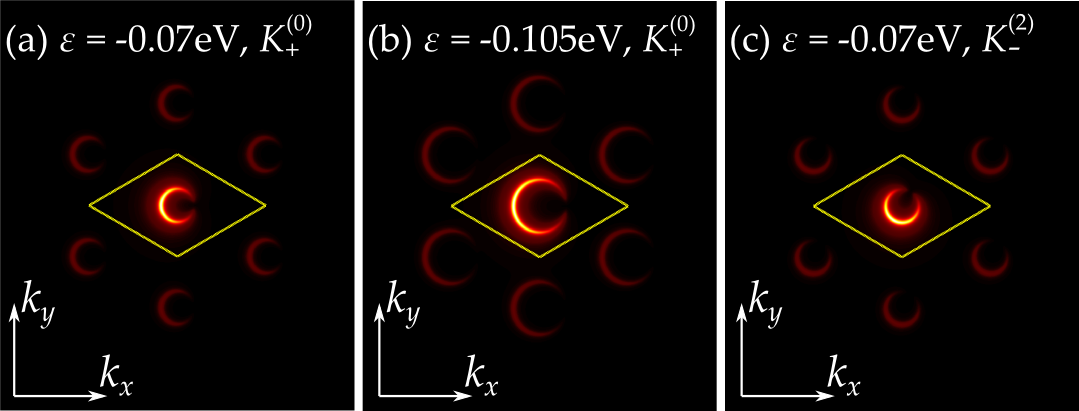}
\caption{ARPES constant-energy maps in the vicinity of valley (a),(b) $\vect{K}_{+}^{(0)}$ and (c) $\vect{K}_{-}^{(2)}$, showing the replicas of the main Dirac cone states due to the moir\'{e}-periodic strain. The yellow rhombus in each panel shows the sBZ boundary and the dimensions of the maps correspond to those of the inset in the right of Fig.~\ref{fig:introduction}.}
\label{fig:strain}
\end{figure}
Because $\vect{p}$ and $\vect{p}_{e}$ are in the vicinity of the same valley, the sums over lattice vectors $\vect{R}$ yield Dirac delta functions so that the transition amplitude is
\begin{align}\label{eqn:strain}
& ^{s}\zeta^{\vect{K}_{+}+\vect{p}_{e}}_{\vect{K}_{+}+\vect{p}} = \left[ \delta(\vect{p}-\vect{p}_{e}) + i\sum_{n}\delta(\vect{p}-\vect{p}_{e}+\vect{b}_{n})(\vect{p}-\vect{p}_{e})\cdot\vect{u}_{n} \right] \nonumber \\ & \times \left[ e^{i(\vect{p}-\vect{p}_{e})\cdot\vect{\tau}_{A}} + se^{i\varphi_{\vect{p}}}e^{i(\vect{p}-\vect{p}_{e})\cdot\vect{\tau}_{B}} \right].
\end{align}
and the intensity
\begin{widetext}
\begin{align}\begin{split}
I(\vect{K}_{\xi}^{(0)}\!+\!\vect{p}_{e}) & \sim \sum_{s} \left( \left| 1+\xi se^{i\varphi_{\vect{p}_{e}}} \right|^{2} + \left| \sum_{n}\vect{b}_{n}\cdot\vect{u}_{n} \left[ e^{-i\vect{b}_{n}\cdot\vect{\tau}_{A}}+\xi se^{i\varphi_{\vect{p}_{e}-\vect{b}_{n}}}e^{-i\vect{b}_{n}\cdot\vect{\tau}_{B}} \right] \right|^{2} \right)\delta(\epsilon_{e}+W-\epsilon^{s}_{+,\vect{p}_{e}}-\omega) \\
& \approx \sum_{s} \left( \left| 1+\xi se^{i\varphi_{\vect{p}_{e}}} \right|^{2} + \left| \sum_{n}\vect{b}_{n}\cdot\vect{u}_{n} \left[ 1+\xi se^{i\varphi_{\vect{p}_{e}-\vect{b}_{n}}} \right] \right|^{2}\right)\delta(\epsilon_{e}+W-\epsilon^{s}_{+,\vect{p}_{e}}-\omega),
\end{split}\end{align}
\end{widetext}
where in the second line we used the fact that $\vect{b}_{n}\cdot\vect{\tau}_{i}\sim\delta\tfrac{2\pi}{3}\ll 1$. Note that transverse deformations, $\vect{b}_{n}\perp\vect{u}_{n}$, do not contribute to the second term in the round bracket above. 

For BZ corners other then $\vect{K}_{\xi}^{(0)}$, the phase factors in the sum over $\vect{R}$ in Eq.~\eqref{eqn:strain1} contain an additional graphene reciprocal vector $[\vect{K}_{\xi}^{(0)}\!-\!\vect{K}_{\xi}^{(j)}]$. However, we can choose any two inequivalent valleys to construct wave functions in Eq.~\eqref{eqn:unperturbed} and Hamiltonian in Eq.~\eqref{eqn:dirac_hamiltonian}. For those two new `reference' valleys, we can follow the same procedure as outlined above for $\vect{K}_{\xi}^{(0)}$, although the coefficients $\chi_{i,\xi}^{s}$ gain additional phase factors. \cite{bena_njp_2009} The general form of the ARPES intensity which preserves the rotational relation between ARPES maps at various BZ corners is then
\begin{align}\begin{split}
& I(\vect{K}^{(j)}_{\xi}\!+\!\vect{p}_{e}) \sim \sum_{s} \left( \left| 1+\xi se^{i\varphi_{\vect{p}_{e}}}e^{i\xi\vect{K}_{\xi}^{(j)}\cdot\vect{d}} \right|^{2} \right. \\ & \left. + \left| \sum_{n}\vect{b}_{n}\cdot\vect{u}_{n} \left[ 1+\xi se^{i\varphi_{\vect{p}_{e}-\vect{b}_{n}}}e^{i\xi\vect{K}_{\xi}^{(j)}\cdot\vect{d}} \right] \right|^{2}\right)\\ & \times \delta(\epsilon_{e}+W-\epsilon^{s}_{+,\vect{p}_{e}}-\omega).
\end{split}\end{align}

We use $\vect{b}_{n}\cdot\vect{u}_{n} = 4\pi\delta$ to plot the ARPES maps in the vicinity of $\vect{K}_{+}^{(0)}$ and $\vect{K}_{-}^{(2)}$ shown in Fig.~\ref{fig:strain}. The moir\'{e}-periodic strain generates satellite peaks shifted by vectors $\vect{b}_{n}$ from the centre of the valley. This additional six-fold structure is clearly distinguishable from the images produced by minibands, Figs.~\ref{fig:i}--\ref{fig:v}(a)--(f). In fact, for ARPES intensity maps for energies close to the Dirac point such replicas of the crescent-shaped image of chiral Dirac electrons can be used to identify the amplitude of strain in graphene because the miniband formation effects are weak in the centre of the sBZ. 

\section{Secondary scattering of graphene photoelectrons by \lowercase{{\it h}}BN}

\begin{figure}[t]
\centering
\includegraphics[width=\columnwidth]{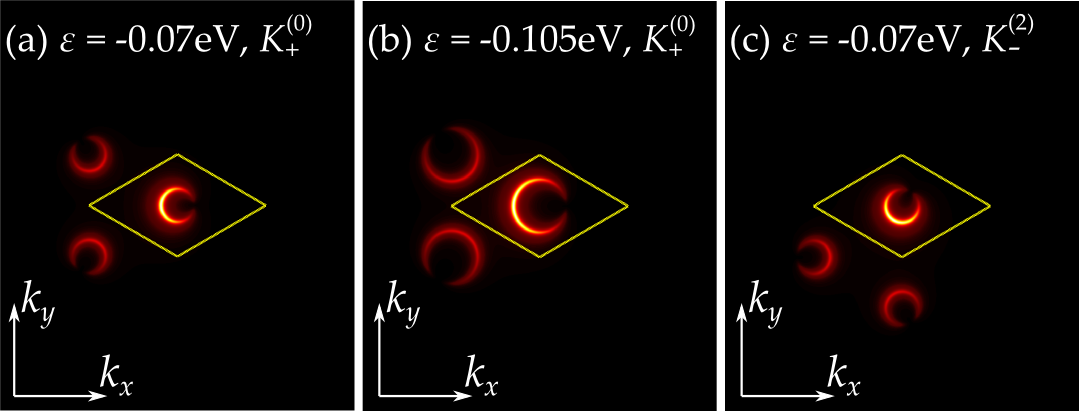}
\caption{ARPES constant-energy maps in the vicinity of valley (a),(b) $\vect{K}_{+}^{(0)}$ and (c) $\vect{K}_{-}^{(2)}$, showing the replicas of the main Dirac cone states due to the scattering of graphene photoelectrons off {\hbn}. The yellow rhombus in each panel shows the sBZ boundary and the dimensions of the maps correspond to those of the inset in the right of Fig.~\ref{fig:introduction}.}
\label{fig:scattering}
\end{figure}

Scattering of the electrons photoemitted from graphene off the underlying {\hbn} before detection changes their momentum by a reciprocal vector of {\hbn} without destroying their memory of the original Dirac state they occupied. We take into account such processes by considering transition amplitude of the form 
\begin{widetext}\begin{align}\begin{split}
& ^{s}\zeta^{\vect{K}_{\xi}^{(j)}+\vect{p}_{e}}_{\vect{K}_{\xi}+\vect{p}} = \int d\vect{q}_{e}\braket{e^{i(\vect{K}_{\xi}^{(j)}+\vect{p}_{e})\cdot\vect{r}}|\sum_{\vect{G}_{\mathrm{BN}}}\alpha_{\vect{G}_{\mathrm{BN}}}e^{i\vect{G}_{\mathrm{BN}}\cdot\vect{r}}|e^{i(\vect{K}_{\xi}^{(j')}+\vect{q}_{e})\cdot\vect{r}}} \braket{e^{i(\vect{K}_{\xi}^{(j')}+\vect{q}_{e})\cdot\vect{r}}e^{iq_{e}^{\perp}z}|\xi,\vect{p},s}_{0} \approx \\
& \sum_{\vect{G},\vect{G}_{\mathrm{BN}}}\!\alpha_{\vect{G}_{\mathrm{BN}}}\hat{\phi}(|\vect{K}_{\xi}^{(j)}\!-\!\vect{G}|,q_{e}^{\perp})\delta(\vect{K}_{\xi}\!+\!\vect{p}\!-\!\vect{K}_{\xi}^{(j)}\!-\!\vect{p}_{e}\!+\!\vect{G}_{\mathrm{BN}}\!-\!\vect{G}) \left[\xi\chi_{A,\xi}^{s}(\vect{p})e^{i\vect{G}\cdot\vect{\tau}_{A}}+\chi_{B,\xi}^{s}(\vect{p})e^{i\vect{G}\cdot\vect{\tau}_{B}}\right],
\end{split}\end{align}\end{widetext}
where $\vect{K}_{\xi}^{(j')}+\vect{q}_{e}$ is the momentum of the photoelectron after emission from graphene and $\vect{K}_{\xi}^{(j)}+\vect{p}_{e}$ is its final momentum after scattering off {\hbn} and gaining additional momentum $\vect{G}_{\mathrm{BN}}$, a reciprocal vector of {\hbn}. The coefficients $\alpha_{\vect{G}_{\mathrm{BN}}}$ characterise efficiency of the scattering off {\hbn} by $\vect{G}_{\mathrm{BN}}$, which we assume depends only on the magnitude of this vector and can also include additional phase shifts due to additional path scattered electrons have to traverse between graphene and {\hbn}. Because the Fourier transform of the $2p_{z}$ orbital $\hat{\phi}(|\vect{p}|,p_{z})$ decays rapidly with increasing $|\vect{p}|$, for each BZ corner three vectors $\vect{G}_{m}$, $m=0,1,2$, for which $|\vect{K}_{\xi}^{(j)}\!-\!\vect{G}_{m}|=|\vect{K}_{+}|=|\vect{K}_{-}|$ provide the greatest contributions to $^{s}\zeta^{\vect{K}_{\xi}^{(j)}+\vect{p}_{e}}_{\vect{K}_{\xi}+\vect{p}}$ [for the valley $\vect{K}_{+}$, for example, they are $\vect{G}_{0}=0$, $\vect{G}_{1}=(\tfrac{2\pi}{a},-\tfrac{2\pi}{a\sqrt{3}})$ and $\vect{G}_{2}=(\tfrac{2\pi}{a},\tfrac{2\pi}{a\sqrt{3}})$]. For those three vectors, we introduce coefficients $\alpha_{0}$ (corresponding to $\vect{G}_{0}=0$ which is always one of those three vectors) and $\alpha_{1}$ (for the other two vectors) and after limiting the sum over $\vect{G}$ to the three biggest terms, we obtain intensity
\begin{widetext}\begin{align}\label{eqn:scattering}\begin{split}
I(\vect{K}_{+}^{(j)}+\vect{p}_{e}) & \sim \sum_{s}\left\{ \left|[1+\alpha_{0}]\left[1+\xi se^{i\varphi_{\vect{p}_{e}}}e^{i(\vect{K}_{+}^{(j)}-\vect{K}_{+}^{(0)})\cdot\vect{d}}\right]\right|^{2} + \left|\alpha_{1}\left[ 1+\xi se^{i\varphi_{\vect{p}_{e}-\xi\op{R}_{2\pi j/3}\vect{b}_{1}}}e^{i(\vect{K}_{+}^{(j)}-\vect{K}_{+}^{(1)})\cdot\vect{d}} \right]\right|^{2} \right.\\ &+ \left. \left|\alpha_{1}\left[ 1+\xi se^{i\varphi_{\vect{p}_{e}-\xi\op{R}_{2\pi j/3}\vect{b}_{2}}}e^{i(\vect{K}_{+}^{(j)}-\vect{K}_{+}^{(2)})\cdot\vect{d}} \right]\right|^{2} \right\} \delta(\epsilon_{e}\!+\!W\!-\!\epsilon^{s}_{+,\vect{p}_{e}}\!-\!\omega).
\end{split}\end{align}\end{widetext}
Here, we also included the contribution of electrons photoemitted from unperturbed graphene that travelled directly to the detector. The first term in Eq.~\eqref{eqn:scattering} has the form identical to the contribution from unperturbed graphene, whereas the second and third term describe contributions from photoelectrons ejected from the vicinity of the BZ corners at $\vect{K}_{\xi}^{(j)}\!+\!\vect{G}_{1}$ and $\vect{K}_{\xi}^{(j)}\!+\!\vect{G}_{2}$, respectively, which scatter off {\hbn} with addition of nonzero $\vect{G}_{\mathrm{BN}}$ and hence are detected at $\vect{K}_{\xi}^{(j)}\!+\!\vect{p}_{e}$. The $A$ and $B$ sublattice components remember the original BZ corner of the electron, so that these two terms generate rotated crescent shapes, as shown in Fig.~\ref{fig:scattering} for the vicinity of $\vect{K}_{+}^{(0)}$ and $\vect{K}_{-}^{(2)}$. To obtain those ARPES maps, we used real $\alpha_{0}$ and $\alpha_{1}$ and $\tfrac{\alpha_{1}}{1+\alpha_{0}}=\tfrac{1}{4}$. Importantly, the additional patterns can be distinguished from the miniband effects by their angular orientation. 

\section{Summary}

To summarise, we show how ARPES can be used to characterise the electronic minibands formed in graphene due to the moir\'{e} potential and hence elucidate on the microscopic details of {\ghbn} heterostructures. We also discuss how the features due to miniband formation can be distinguished from those due to the periodic strain in graphene. We note that nonzero angular misalignment $\theta$ between graphene and {\hbn} changes the size of and rotates the sBZ but does not affect the angular orientation of the additional crescent shapes appearing due to the moir\'{e}-periodic strain or photoelectron scattering off the substrate (because these are replicas of crescent patterns formed by Dirac states around BZ corners related by graphene reciprocal vector).

\begin{acknowledgements}
The authors would like to acknowledge stimulating discussions with Shuyun Zhou and Eryin Wang. This work has been supported by the EPSRC First Grant EP/L013010/1 (MM-K), FP7 European Graphene Flagship Project CNECT-ICT-213604391 (JRW) and ERC (Belgium) Synergy Grant Hetero2D (VIF). MM-K acknowledges the hospitality of the NUS Graphene Research Centre.
\end{acknowledgements}


\begin{thebibliography}{99}
\bibitem{strain_footnote} In fact, we take as $\vect{u}(\vect{r})$ only the periodic part of the deformation. This is equivalent to the strain present in the lattice if the moir\'{e} unit cell does not change its area as a result of the deformation. If this is not the case, a homogeneous hydrostatic expansion/contraction should be added to $\vect{u}(\vect{r})$ to obtain the true strain field. The hydrostatic part, however, has no effect on the ARPES pattern beyond rescaling the momentum scale $b$.
\end{thebibliography}
\end{document}